\documentclass[10pt]{article}
\usepackage{fullpage}
\usepackage{amsmath}
\usepackage{amssymb}
\usepackage[dvips]{epsfig}
\usepackage{color}

\def\##1{\underline{#1}}
\def\=#1{\underline{\underline{#1}}}

\def\+#1{\underline{\bf #1}}
\def\*#1{\underline{\underline{\bf #1}}}

\def\r#1{(\ref{#1})}
\def\l#1{\label{#1}}
\def\c#1{\cite{#1}}

\def\le{\left(}
\def\ri{\right)}
\def\les{\left[}
\def\ris{\right]}
\def\lec{\left\{}
\def\ric{\right\}}

\def\.{\mbox{ \tiny{$^\bullet$} }}

\def\epso{\epsilon_{\scriptscriptstyle 0}}

\def\muo{\mu_{\scriptscriptstyle 0}}

\begin{document}

\begin{center}

{\bf {\LARGE Towards a metamaterial simulation \vspace{8pt} \\of a
spinning cosmic string }}

\vspace{10mm} \large

 Tom G. Mackay\footnote{E--mail: T.Mackay@ed.ac.uk.}\\
{\em School of Mathematics and
   Maxwell Institute for Mathematical Sciences\\
University of Edinburgh, Edinburgh EH9 3JZ, UK}\\
and\\
 {\em NanoMM~---~Nanoengineered Metamaterials Group\\ Department of Engineering Science and Mechanics\\
Pennsylvania State University, University Park, PA 16802--6812,
USA}\\
 \vspace{3mm}
 Akhlesh  Lakhtakia\footnote{E--mail: akhlesh@psu.edu}\\
 {\em NanoMM~---~Nanoengineered Metamaterials Group\\ Department of Engineering Science and Mechanics\\
Pennsylvania State University, University Park, PA 16802--6812, USA}

\end{center}

\vspace{4mm}

\normalsize

\begin{abstract}

Establishing the constitutive parameters of a nonhomogeneous
bianisotropic medium that is  equivalent to the spacetime metric of
a spinning cosmic string, in a noncovariant formalism, we  found   a
metamaterial route to investigate the existence of closed timelike
curves.

\end{abstract}

\vspace{5mm} \noindent  PACS numbers: 04.40.Nr, 41.20.-q, 98.80.Cq

\section{Introduction}

Metamaterials are artificial composite materials which---through
judicious design---exhibit remarkable characteristics not displayed
by their constituent materials \c{Walser}.
Most conspicuously, recent research efforts have been directed
towards the development of metamaterials which support negative
refraction \c{SAR} and those which facilitate a degree of
concealment \c{Greenleaf}. Metamaterials may also provide useful
models for the study of general relativistic scenarios which are
otherwise impractical or impossible to explore
\c{Genov,de_Sitter_metamaterial,Black_hole_metamaterial,Big_bang_metamaterial}.
Furthermore, certain metrics, such as for the
 Schwarzschild \c{MSL04}, Schwarzschild--de
Sitter \c{MLS_EurophysLett}, Kerr \c{MLS_NJP_Kerr,Kom,Sharif},
Kerr--Newman \c{Kerr_Newman} and Reissner--Nordstr\"om \c{Hossain}
spacetimes, are associated with negative-phase-velocity propagation of light---a
notable property of certain
negatively refracting metamaterials \c{ML_PRB}---in a noncovariant formalism.

The formal analogy between light propagation in vacuum subjected to
a gravitational field and   light propagation in certain
nonhomogeneous bianisotropic mediums in flat spacetime has been
appreciated for at least 85 years \c{Tamm}.  The realization of such bianisotropic
mediums is moving towards practicality, since the
emergence of metamaterials. The aim of this
letter is to establish the constitutive relations of a
nonhomogeneous bianisotropic medium which is formally equivalent to
the spacetime metric associated with a spinning cosmic string
\c{Lorenci}. The metric describing spinning cosmic strings is
particularly interesting because it can support the existence of closed timelike
curves (CTCs) \c{Slobodov}. But CTCs are problematic as they   violate the
principle of causality. Accordingly, a
practical model to simulate this spacetime metric would be of
considerable value.

\section{Spacetime metric of a spinning cosmic string}

Consistently with the signature $\le -, +, +, + \ri$,
 the line element of a spinning cosmic string has the
form \c{Marder}
\begin{equation}
ds^2 = - \le F \, dt + M\, d \phi \ri^2 + A^2 \, d \phi^2 + d z^2 +
d\rho^2, \l{line_el}
\end{equation}
 in  terms of time $t$ and cylindrical
polar coordinates $\le \rho, \phi, z \ri$. We adopt the
``ballpoint pen" model of Jensen and Soleng \c{Jensen_Soleng}
wherein the metric describing the interior region of the string
$\rho < \rho_s$ is matched smoothly at $\rho=\rho_s$ to the metric describing the
string's exterior region $\rho > \rho_s$. Accordingly, we have $F
\equiv 1$ with
\begin{equation}
A =  \left\{
\begin{array}{lr}
\displaystyle{\frac{1}{\sqrt{\lambda}} \, \sin \le \rho
\sqrt{\lambda}
 \ri,} & \rho \leq \rho_s \vspace{6pt}\\ \displaystyle{\le 1 - 4 G \nu \ri
\les \rho + \rho_s \le \frac{\tan \le \rho_s \sqrt{\lambda} \ri
}{\rho_s \sqrt{\lambda}} -1 \ri \ris,} & \rho > \rho_s
\end{array}
\right.
\end{equation}
and
\begin{equation}
M =  \left\{
\begin{array}{lr}
\displaystyle{2 \alpha \les \le \rho - \rho_s \ri \cos \le \rho
\sqrt{\lambda}
 \ri  - \frac{1}{\sqrt{\lambda}} \, \sin \le
\rho \sqrt{\lambda}  \ri + \rho_s \ris,} & \rho \leq \rho_s \vspace{6pt}\\
\displaystyle{ 4 G J,} & \rho
> \rho_s
\end{array}.
\right.
\end{equation}
 Herein, $\nu$ and $J$
represent the mass  and angular momentum per unit length of the
string, respectively; $G$ is the gravitational constant; $\lambda $
is a positive--valued constant; and $\alpha \leq 1$. Smooth matching
at $\rho = \rho_s$ imposes the equalities
\begin{equation}
\left.
\begin{array}{l}
\nu = \displaystyle{ \frac{1}{4 G} \les 1 - \cos \le \rho_s
\sqrt{\lambda} \ri \ris}
\vspace{4pt}\\
J = \displaystyle{ \frac{\alpha}{2 G}  \les  -
\frac{1}{\sqrt{\lambda}} \, \sin \le \rho_s \sqrt{\lambda}  \ri +
\rho_s \ris}
\end{array}
\right\}.
\end{equation}

After replacing the cylindrical polar coordinates by the Cartesian coordinates $\le  x, y, z \ri$, the
spacetime metric is  represented by  $g_{\alpha \beta}$ per
\footnote{Roman indexes take the values 1, 2 and 3, while Greek
indexes take the values 0, 1, 2 and 3.}
\begin{eqnarray}
 \les \, g_{\alpha \beta} \, \ris & = &  \left(
\begin{array}{cccc}
-1 & \displaystyle{ \frac{M y}{\rho^2} }& \displaystyle{ - \frac{M x}{\rho^2}} & 0\\
\\
\displaystyle{ \frac{M y}{\rho^2}} &\displaystyle{ \frac{\rho^2 x^2 +
\le A^2 - M^2 \ri y^2}{\rho^4}}
 & \displaystyle{ \frac{\le \rho^2 - A^2 + M^2 \ri x y }{\rho^4}} & 0 \\
\\
\displaystyle{- \frac{M x}{\rho^2}} & \displaystyle{\frac{\le \rho^2
- A^2 + M^2 \ri x y }{\rho^4}} & \displaystyle{ \frac{\rho^2 y^2 +
\le A^2 - M^2 \ri x^2}{\rho^4} }
 & 0 \\
\\
0 & 0 & 0 &
 1
\end{array} \right), \l{gab}
\end{eqnarray}
where $\rho^2 = x^2 + y^2$. We note the following limits:
\begin{eqnarray}
\lim_{\rho \to \infty} \les \, g_{\alpha \beta} \, \ris & = &
\left(
\begin{array}{cccc}
-1 & 0 & 0 & 0\\
\\
0 & \cos^2  \phi  + \cos^2 \le \rho_s \sqrt{\lambda} \ri \, \sin^2
\phi
 &  \sin^2 \le \rho_s \sqrt{\lambda} \ri \, \cos  \phi  \, \sin \phi & 0 \\
\\
0 & \sin^2 \le \rho_s \sqrt{\lambda} \ri \, \cos  \phi  \, \sin \phi
& \sin^2  \phi  + \cos^2 \le \rho_s \sqrt{\lambda} \ri \, \cos^2
\phi
 & 0 \\
\\
0 & 0 & 0 &
 1
\end{array} \right), \nonumber \\ &&  \\
\lim_{\rho \to 0} \les \, g_{\alpha \beta} \, \ris & = & \mbox{diag}
\, \le -1, 1, 1, 1 \ri,
\end{eqnarray}
where $\cos \phi = x/\rho$ and $\sin \phi = y/\rho$. In particular, in the limit $\rho \to \infty$,
 the eigenvalues of $  \les \, g_{\alpha \beta} \,
\ris$ are $\lec -1, \, 1, \,  1,\, \cos^2 \le
\rho_s \sqrt{\lambda} \ri \ric$. Accordingly, $  \les \, g_{\alpha
\beta} \, \ris \to \mbox{diag} \, \le -1, 1, 1, 1 \ri$ as $\rho \to
\infty$ only when $\rho_s \sqrt{\lambda}$ is an integer multiple of
$\pi$.

Following the approach of Tamm \c{Tamm}---which was later
developed  by others \c{Skrotskii,Plebanski,SS,Mashhoon}---electromagnetic
fields in the curved spacetime associated with a
spinning cosmic string may be described by the constitutive
relations of an equivalent medium per
\begin{equation}
\label{CR2} \left.
\begin{array}{l}
D_\ell = \epso \gamma_{\ell m} E_m +  \sqrt{\epso \muo} \, \varepsilon_{\ell mn}\,\Gamma_m\,H_n\\[6pt]
B_\ell =  \muo  \gamma_{\ell m} H_m - \sqrt{\epso \muo} \,
\varepsilon_{\ell mn}\, \Gamma_m\, E_n
\end{array}\right\},
\end{equation}
in SI units. Herein, the scalar constants $\epso$ and $\muo$ denote
the permittivity and permeability of vacuum in the absence of a
gravitational field;
 $\varepsilon_{\ell mn}$ is the three--dimensional
Levi--Civita symbol; and the components of  $\gamma_{\ell m}$ and
$\Gamma_m$  are defined as
\begin{equation}
\label{akh1} \left.\begin{array}{l} \gamma_{\ell m}
= \displaystyle{ \sqrt{ -g}  \, \frac{{g}^{\ell m}}{{g}_{00}}}\\[6pt]
\Gamma_m= \displaystyle{\frac{g_{0m}}{g_{00}}}
\end{array}\ric,
\end{equation}
where the sign of the square root term in the definition of
$\gamma_{\ell m}$ is chosen to ensure that the metric for
Minkowskian spacetime is represented by the matrix $\les
\gamma_{\ell m} \ris = \mbox{diag} \, \le 1,\, 1, \, 1 \ri$.
 Thus, in this noncovariant formalism, the curved spacetime
associated with a spinning cosmic string is represented by the fictitious, nonhomogeneous,
spatiotemporally local, bianisotropic medium
characterized by \r{CR2}. With respect to the Cartesian basis
vectors, the matrix and vector representations of $\gamma_{\ell m}$
and $\Gamma_m$ are
\begin{equation}
\les \gamma_{\ell m} \ris = \frac{1}{\rho} \le
\begin{array}{ccc}
\displaystyle{\frac{A^2 x^2 + \rho^2 y^2}{A \rho^2}} &
\displaystyle{\frac{\le A^2 - \rho^2 \ri x y }{A \rho^2}} & 0
\\ \\
\displaystyle{\frac{\le A^2 - \rho^2 \ri x y }{A \rho^2}} &
\displaystyle{\frac{A^2 y^2 + \rho^2 x^2}{A \rho^2}} &0 \\ \\ 0 & 0
& A
\end{array}
 \ri \l{matrix_gamma}
\end{equation}
and
\begin{equation}
\les \Gamma_{ m} \ris = \frac{M}{\rho^2}  \le -y, \, x, \, 0 \ri.
\l{vector_gamma}
\end{equation}

We note the following limits:
\begin{eqnarray}
&& \lim_{\rho \to \infty} \les \gamma_{\ell m} \ris = \nonumber  \\
&& \sqrt{ \cos^2 \le \rho_s \sqrt{\lambda} \ri }
 \le
\begin{array}{ccc}
\displaystyle{ \cos^2 \phi + \frac{ \sin^2 \phi}{ \cos^2 \le \rho_s
\sqrt{\lambda} \ri  }} & - \displaystyle{ \tan^2 \le \rho_s
\sqrt{\lambda} \ri \,  \cos \phi \sin \phi} & 0 \\ \\
- \displaystyle{\tan^2 \le \rho_s \sqrt{\lambda} \ri \,  \cos \phi
\sin \phi} & \displaystyle{ \sin^2 \phi + \frac{ \cos^2 \phi}{
\cos^2
\le \rho_s \sqrt{\lambda} \ri  }} &0 \\ \\
 \\ \\ 0 & 0 & 1
\end{array}
 \ri, \nonumber \\ &&  \\
&& \lim_{\rho \to 0} \les \gamma_{\ell m} \ris = \mbox{diag} \, \le
1,\, 1, \, 1 \ri\,,
\\[5pt]
&& \lim_{\rho \to \infty} \les \Gamma_{ m} \ris = \lim_{\rho \to 0}
\les \Gamma_{ m} \ris
 = \mbox{diag} \, \le 0,\, 0, \, 0 \ri.
\end{eqnarray}
In particular, we observe that $  \les \, \gamma_{\ell m} \, \ris
\to \mbox{diag} \, \le  1, 1, 1 \ri$ as $\rho \to \infty$ only when
$\rho_s \sqrt{\lambda}$ is an integer multiple of $ \pi$.

\section{Numerical illustration}
\begin{figure}[!ht]
\centering \psfull \epsfig{file=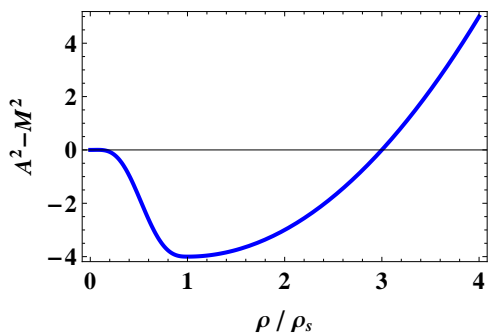,width=2.7in}
\caption{The quantity $A^2 - M^2$
  plotted versus $\rho / \rho_s$, for $\alpha =1$ and  $r_s \sqrt{\lambda}
=  \pi$. } \label{fig1}
\end{figure}

Following Jensen and Soleng \c{Jensen_Soleng}, we take $\alpha = 1$.
Let  us present numerical results for the limiting case where $\nu$
attains its maximum value of $1/\le 2 G \ri$. Closed timelike curves
are supported when the coefficient $(A^2 - M^2)$ of $d \phi^2$ in
the line--element expression \r{line_el} is negative--valued
\c{Slobodov}. This quantity is plotted against $\rho / \rho_s$ in
Fig.~\ref{fig1}. We find that $A^2 - M^2 < 0 $ for $0.11 < \le \rho
/ \rho_s \ri < 3$; i.e., both the interior and exterior string
regions support CTCs.

\begin{figure}[!h]
\centering \psfull \epsfig{file=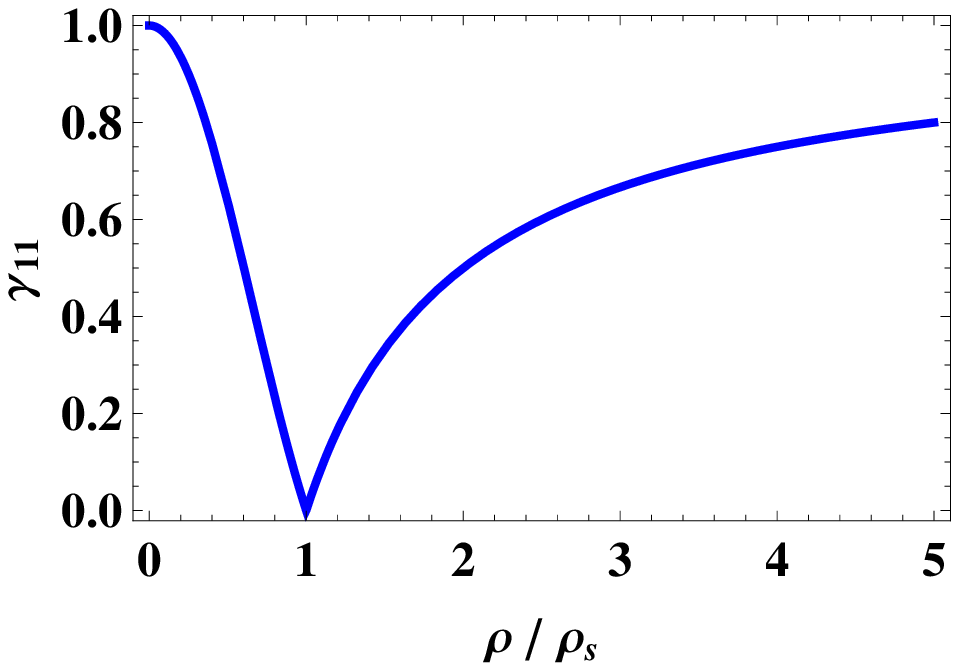,width=2.7in}
\epsfig{file=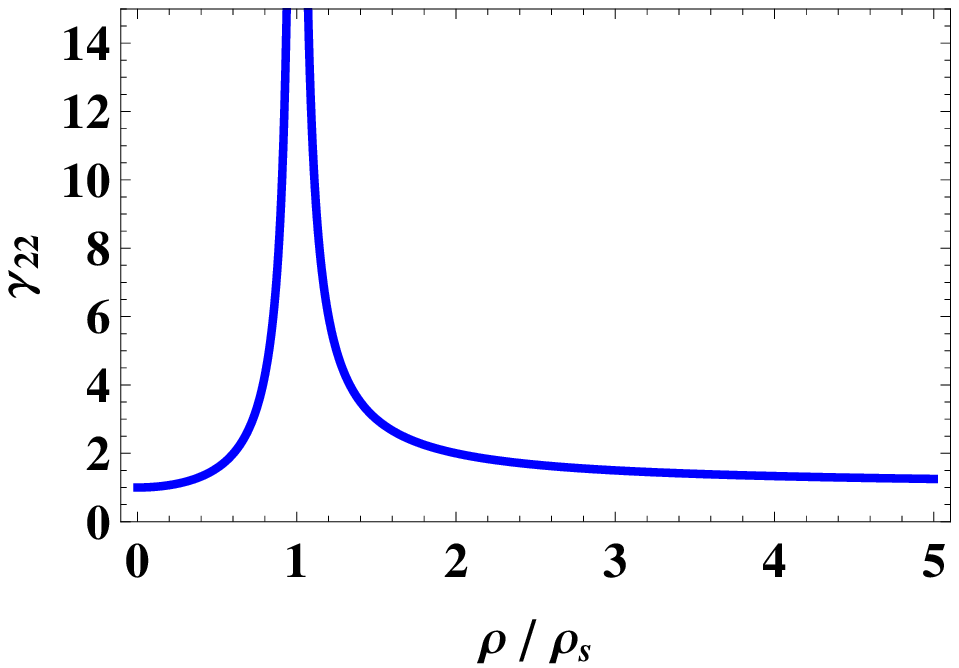,width=2.7in}
 \epsfig{file=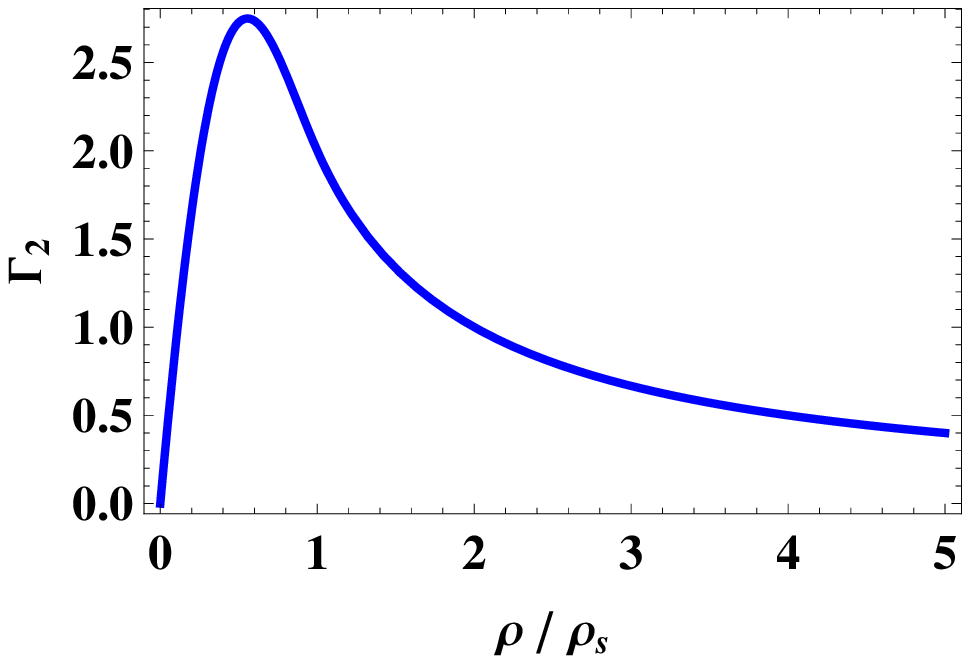,width=2.7in}
 \caption{The constitutive quantities $\gamma_{11} $, $\gamma_{22}$ and
$\Gamma_{2}$ plotted versus $\rho / \rho_s$, for $\alpha =1$ and
$r_s \sqrt{\lambda} =  \pi$. } \label{fig2}
\end{figure}

The fictitious bianisotropic medium characterized by \r{CR2} becomes
equivalent to vacuum (in flat spacetime) in the limits $\rho \to 0$
and $\rho \to \infty$, for all values of $\phi$. For definiteness,
let us  fix the angle $\phi = 0$. The matrix $ \les \, \gamma_{\ell
m} \, \ris $ then is diagonal with $\gamma_{11} \equiv
\gamma_{33}$, and the vector $ \les \, \Gamma_{ m} \, \ris $ has
only one nonzero component: $\Gamma_2 = M / \rho$. The
constitutive parameters $\gamma_{11}$, $\gamma_{22}$ and $\Gamma_2$
are plotted against $\rho / \rho_s$ in Fig.~\ref{fig2}. The
constitutive parameters $\gamma_{11}$ and $\Gamma_2$ remain bounded
for all values of $\rho / \rho_s$, whereas $\gamma_{22}$ becomes
unbounded at $\rho = \rho_s$. For a metamaterial simulation of a cosmic string, one would have to use
$\gamma_{22}\gg1$  at $\rho = \rho_s$

The limiting behavior displayed in Fig.~\ref{fig2} as $\rho \to 0$
and $\rho \to \infty$ also arises when $\nu$ attains its minimum
value of zero. Therefore, a realistic metamaterial simulation of a
massless string \c{Culetu} is also possible. Just as for $\nu = 1 /
\le 2 G \ri $, our calculations (not presented here) reveal that
$\nu = 0$ is also compatible with the existence of CTCs in both the
interior and exterior regions.

\section{Closing remarks}
The chief purpose of this letter was to present the constitutive
matrix \r{matrix_gamma} and vector \r{vector_gamma} which together specify
the fictitious bianisotropic medium that simulates the   curved spacetime associated with a
spinning cosmic string. Thereby, a recipe for  a simulating
metamaterial has been formulated. Some variation of this recipe could be implemented to
resolve fundamental issues pertaining to the existence of CTCs.

\vspace{10mm}

\noindent {\bf Acknowledgments:}
 TGM is supported by a  Royal
Academy of Engineering/Leverhulme Trust Senior Research Fellowship.
AL thanks the Binder Endowment at Penn State for partial financial
support of his research activities.

\end{document}